\documentclass[12pt]{article}
\usepackage{amsmath, amssymb,graphicx}

\newwrite\ffile\global\newcount\figno \global\figno=1

\def\writedef#1{}

\input epsf
\def\figin{\epsfcheck\figin}\def\figins{\epsfcheck\figins}
\def\epsfcheck{\ifx\epsfbox\UnDeFiNeD
\message{(NO epsf.tex, FIGURES WILL BE IGNORED)}
\gdef\figin##1{\vskip2in}\gdef\figins##1{\hskip.5in}
\else\message{(FIGURES WILL BE INCLUDED)}%
\gdef\figin##1{##1}\gdef\figins##1{##1}\fi}

\def\figinsert{}
\def\ifig#1#2#3{\xdef#1{fig.~\the\figno}
\writedef{#1\leftbracket fig.\noexpand~\the\figno}%
\figinsert\figin{\centerline{#3}}\medskip\centerline{\vbox{\baselineskip12pt
\advance\hsize by -1truein\center\footnotesize{  Fig.~\the\figno.} #2}}
\bigskip\endinsert\global\advance\figno by1}
\def\endinsert{}

\usepackage{amssymb}
\usepackage{graphics}
\usepackage{subfigure}
\begin{document}
\baselineskip 18pt
\newcommand{\Tr}{\mbox{Tr\,}}
\newcommand{\beq}{\begin{equation}}
\newcommand{\eeq}{\end{equation}}
\newcommand{\bea}{\begin{eqnarray}}
\newcommand{\eea}[1]{\label{#1}\end{eqnarray}}
\renewcommand{\Re}{\mbox{Re}\,}
\renewcommand{\Im}{\mbox{Im}\,}
\newcommand{\yms}{${YM^*\,}$}

\def\N{{\cal N}}
\def\d{{\rm d\,}\!}


\thispagestyle{empty}
\renewcommand{\thefootnote}{\fnsymbol{footnote}}

{\hfill \parbox{4cm}{
        SHEP-05-05 \\
}}

\bigskip

\begin{center} \noindent \Large \bf
D7 Brane Embeddings and Chiral Symmetry Breaking
\end{center}

\bigskip\bigskip\bigskip

\centerline{ \normalsize \bf Nick Evans, Jon Shock, Tom Waterson
\footnote[1]{\noindent \tt
 evans@phys.soton.ac.uk, jps@hep.phys.soton.ac.uk, trw@hep.phys.soton.ac.uk} }

\bigskip
\bigskip\bigskip

\centerline{ \it Department of Physics} \centerline{ \it
Southampton University} \centerline{\it  Southampton, SO17 1BJ }
\centerline{ \it United Kingdom}
\bigskip

\bigskip\bigskip

\renewcommand{\thefootnote}{\arabic{footnote}}

\centerline{\bf \small Abstract}
\medskip

{\small \noindent We study the embedding of D7 brane probes in
five geometries that are deformations of $AdS_5 \times S^5$. Each
case corresponds to the inclusion of quark fields in a dual gauge
theory where we are interested in investigating whether chiral
symmetry breaking occurs. We use a supersymmetric geometry
describing an ${\cal N}=2$ theory on its moduli space and a
dilaton driven non-supersymmetric flow to establish criteria for a
chiral symmetry breaking embedding. We develop a simple spherical
D7 embedding that tests the repulsion of the core of the geometry
and signals dynamical symmetry breaking. We then use this tool in
more complicated geometries to show that an ${\cal N}=2^*$ theory
and a non-supersymmetric theory with scalar masses do not induce a
chiral condensate. Finally we provide evidence that the Yang
Mills$^*$ geometry does.}

\newpage


\section{Introduction}

A pleasing geometric picture of chiral symmetry breaking in
strongly coupled gauge theory has emerged from dual brane
constructions
\cite{Babington:2003vm}\cite{Kruczenski:2003uq}\cite{GY:2004}-\cite{Armoni:2004}.
Quark fields can be introduced into the AdS/CFT Correspondence
\cite{Mal}-\cite{Wit} by the introduction of new branes
\cite{KarchKatz}\cite{Weiner}\cite{MateosMyers}-\cite{Sakai:2004};
strings stretched between these branes and the branes of the
original construction transform in the fundamental representations
of a product gauge group. The scalar field describing the
separation of these two types of branes is dual to the quark mass
and quark condensate operator $\bar{\psi}\psi$. The embedding of
the new branes in the space created by the original therefore
provides a dynamical determination of whether there is a fermion
bilinear operator that breaks chiral symmetry. Such a symmetry is
normally a manifest symmetry of the spacetime, and the brane's
embedding can be explicitly seen to dynamically break it. These
constructions allow the study of a number of phenomenological
aspects of strongly coupled gauge theories including the pions and
thermal properties \cite{Babington:2003vm}- \cite{Armoni:2004}.

So far this phenomena has been studied in a small number of
geometrically simple spacetimes. The simplest cases consist of
placing a probe D7 brane into a geometry that is a deformation of
$AdS_5 \times S^5$. Asymptotically at large radius or high energy
the theory becomes the ${\cal N}=2$ theory with hypermultiplets of
Karch and Katz \cite{KarchKatz}, which is well understood. When
the deformation is caused by the R-chargeless dilaton switching on
\cite{CM}-\cite{Liu}, the SO(6) symmetry of the space is untouched
and the embedding equations for the probe D7 are easily solved to
give a profile dependent on the radial direction in the space
\cite{Babington:2003vm}\cite{GY:2004}-\cite{Bak:2004}. Similarly,
deformations due to the inclusion of finite temperature
\cite{Babington:2003vm} leave the $S^5$ unspoilt and allow a simple
analysis. Typically such deformations have a singular core to the
geometry indicating the presence of a multi-centre like or fuzzy
D3 brane distribution. In non-supersymmetric configurations, which
can allow the fermion condensate, these are hard to resolve, but in
chiral symmetry breaking cases it is the repulsion of the probe D7
by this core that cause the fermion condensate. We can escape
without a precise knowledge of the singularity.

In \cite{Kruczenski:2003uq} a parallel analysis studies a six
dimensional theory reduced by compactification to a four
dimensional gauge theory in the infra-red (IR). It again has a
repulsive, but this time non-singular, core that induces chiral
symmetry breaking. Again the space-time is sufficiently simple
that the embedding equations may be straightforwardly solved.

It would be nice to have a wider understanding of how generic the
phenomena of chiral symmetry breaking is in gravitational duals -
naively one might expect any non-supersymmetric, strongly coupled
gauge background to induce chiral symmetry breaking. In this paper
we pursue this goal by studying D7 brane embeddings in a wider
range of deformations of $AdS_5 \times S^5$. In general such
geometries are considerably more complicated and the embedding
profile depends on many of the coordinates on the brane. This
turns finding the solutions into an extended and complicated
numerical problem. Instead we propose a simple spherical embedding
of a D7 brane that can be performed analytically and tests the
repulsion of the core of the geometry. We believe this is a good
indication of whether chiral symmetry would be induced. Further,
in the case of a known dilaton induced flow, it provides an
analytic estimate of the dynamical quark mass at zero energy which
matches the numerical results well.

We begin by looking at a deformed geometry of Freedman, Gubser,
Pilch and Warner \cite{FGPW}. In fact, this geometry is a
coordinate transformation of an ${\cal N}=4$ preserving
multi-centre solution in which chiral symmetry breaking can not
occur. Study of it though highlights the pitfalls of interpreting
the numerical embedding solutions too glibly. In one coordinate
system we find solutions for the D7 brane embedding that appear to
give a non-zero value for the quark condensate in the ultra-violet
(UV) when in fact there is none. We learn that for there to be
chiral symmetry breaking there must be a spatial gap between the
embedded D7 brane and the core singularity for the case of a
massless quark in the UV.

We next quickly review the dilaton induced flow results
\cite{Babington:2003vm}\cite{GY:2004}-\cite{Bak:2004} and
introduce our spherical D7 embedding showing how the full
solutions corresponding to the usual D7 embedding match on to
these solutions in the infra-red. The spherical embedding
illustrates the repulsion of the core of the geometry and provides
the analytic result for the induced mass gap in the theory.

As a first new non-supersymmetric geometry we look at an extension
of the FGPW solution \cite{Babington:2002ci} which allows for the inclusion of non-zero
masses for the adjoint scalars of the ${\cal N}=4$ gauge theory as
well as a vacuum expectation value. The gauge theory here has an
unbounded scalar potential but the supersymmetry breaking could
induce chiral symmetry breaking for the probe. We find that this
is a case where chiral symmetry breaking is not present.

Pilch and Warner have constructed a number of geometries
describing the ${\cal N}=2^*$ gauge theory \cite{PW,EJP,BPP}. We
study a D7 embedding that preserves the ${\cal N}=2$ symmetry. The
spherical D7 embedding allows us to check that there is no
repulsion in this case and hence no chiral symmetry breaking as
one expects in a supersymmetric theory.

Finally we look at the non-supersymmetric Yang Mills$^*$ geometry
\cite{Babington:2002qt}. Here the deformation is instigated by a
mass and or condensate for the adjoint fermions of the ${\cal
N}=4$ theory (one would expect these parameters to be linked but
the supergravity geometry provides inconclusive evidence for what
this link is). This supersymmetry breaking induces a mass for the
adjoint scalars too leaving just the Yang Mills field to survive
to the IR. One would expect this theory to break chiral
symmetries. We find using the spherical embedding that generically
the core is repulsive (although there is a line of flows in the
adjoint mass vs condensate plane where chiral symmetry breaking is
absent).

\section{D7 Branes and ${\cal N}=4$ Geometries}\label{Nefour}

The large $N_c$ ${\cal N}=4$ gauge theory at the origin of its
moduli space is dual, via the AdS/CFT Correspondence to
supergravity on $AdS_5\times S^5$ with radius $R$

\beq ds^2 = {u^2\over R^2} dx_{//}^2 + {R^2 \over u^2}
\sum_{i=1}^6 du_i^2, \eeq where $x_{//}$ is the 3+1d plane
parallel to the D3 world volume and $u^2 = \sum_i u_i^2$. Quark
fields may be introduced \cite{KarchKatz} and ${\cal N}=2$
supersymmetry preserved by placing a D7 probe in the $x_{//}$ and
$u_1-u_4$ directions (we write the metric on these four directions
as $d\rho^2 + \rho^2 d\Omega_3^2$). The U(1) symmetry in the
$u_5-u_6$ plane is the geometric realization of the U(1)$_A$ axial
symmetry of the quark fields on the probe. The Dirac Born Infeld
action for the probe, with tension $T_7$, is

\beq S ~=~ - T_7 \int d^8\xi \sqrt{- {\rm det} G_{ab}} ~=~ -T_7
\int d^4x ~ d \rho ~ d \Omega_3 \rho^3 \sqrt{1 + u_5^{'2} +
u_6^{'2}}\label{ADSDBI}, \eeq where the prime indicates a
derivative with respect to $\rho$, $G_{ab}$ is the  pullback of
the metric onto the probe, and we generically use $d^8\xi$ to
indicate integration over the world volume. The resulting equation
of motion for the profile in, for example, the $u_6$ direction is:

\beq { d \over d \rho} \left[ {\rho^3 u_6' \over \sqrt{1 +
u_6^{'2}}} \right] = 0.\eeq The regular solution is $u_6=m$ with
the separation of the D7 from the $\rho$ axis, $m$, representing
the hypermultiplet's mass. Asymptotically at large $u$ there is a
second solution

\beq u_6 = m+ {c \over \rho^2}+\dots .\eeq The parameter $c$
represents the magnitude of the fermionic quark bilinear
$\bar{\psi} \psi$. In pure AdS this solution is singular in the
interior and unphysical \cite{Babington:2003vm}.

Now as a first example of embedding a D7 brane in a deformed
geometry we consider the geometry of Gubser, Freedman, Pilch and
Warner \cite{FGPW}. This geometry was constructed in the 5d
truncation of IIB supergravity on AdS \cite{GRW}-\cite{GRWb} then
lifted to 10d. A scalar field, $\chi$, in the 20 dimensional
representation of SU(4)$_R$ has a non trivial radial profile
corresponding to the scalar operator $\Tr \phi^2$ having a vev of
the form diag$(1,1,1,1,-2,-2)$. This vev preserves the
supersymmetry of the ${\cal N}=4/2$ theory and we would expect to
be able to introduce quarks of any mass via a D7 brane and find no
chiral symmetry breaking. This is indeed the case, as we will
show, but there are some interesting subtleties.\newpage

The 10d geometry is given by:

\beq ds^2 = {X^{1/2} \over \chi} e^{2 A} dx_{//}^2 - {X^{1/2}
\over \chi}\left( du^2 + {R^2 \over \chi^2} \left[ d \theta^2 +
{\sin^2 \theta \over X} d\phi^2 + {\chi^6 \cos^2 \theta \over X} d
\Omega_3^2 \right]\right). \eeq Here $d\Omega_3^2$ is the metric
of a three sphere, $R$ is the asymptotic (large $u$) AdS radius,
and \beq X= \cos^2 \theta + \chi^6 \sin^2 \theta\label{defineX}
.\eeq The fields $\chi$ and $A$ satisfy the differential equations
\beq { d\chi \over du} = { 1 \over 3 R} \left( {1
\over \chi}-\chi^5\right), \hspace{1cm} { dA \over du} = {2 \over 3 R}
\left( { 1 \over \chi^2} + {\chi^4 \over
2}\right),\eeq with solution \beq e^{2 A} = {l^2 \over R^2}
{\chi^4 \over \chi^6-1},\eeq where $l$ is an integration constant.
The solution also has a non-zero four-form potential \beq C_{(4)}
= {e^{4A} X \over g_s \chi^2} dx^0 \wedge dx^1 \wedge dx^2 \wedge
dx^3. \eeq

Note that this solution is singular at the point where $\chi=1$ -
we will see the physical interpretation of this shortly.

It is natural to embed the D7 brane in this geometry to lie in the
$x_{//}, u, \Omega_3$ directions and look at the profile of the D7
brane $\theta(u)$ at fixed $\phi$ ($\phi$ provides the U(1)
symmetry of the embedding which is the chiral symmetry of the
theory with quarks). The DBI action is

\begin{equation}
S_{DBI}=-T_7 \int{d^8\xi}~e^{4A(u)} R^3
|\cos^3\theta(u)|\sqrt{X}\chi(u)^2 \sqrt{1+\frac{\theta'(u)^2
R^2}{\chi(u)^2}}.\label{SUSYDBI}
\end{equation}
To place the solutions in a cartesian like plane instead we can
make a change of coordinates from the circular coordinates
$(u,\theta)$ into the set $(r,v)$:
\begin{eqnarray}
r^2+v^2&=&e^{2u},\nonumber\\
\frac{v}{r}&=&\tan\theta\label{cov}.
\end{eqnarray}
Using this coordinate transformation, the metric becomes:
\begin{eqnarray}
ds^2 = {\sqrt{X} \over \chi} e^{2 A} dx_{//}^2& - &
\frac{\sqrt{X}}{\chi\left(v^2+r^2\right)^2}\left(dv^2\left(v^2+
\frac{R^2 r^2}{\chi^2}\right)+dr^2\left(r^2+\frac{R^2 v^2}{\chi^2}\right)
+2r v dv dr\left(1-\frac{1}{\chi^2}\right)\nonumber\right.\\
&+&\left.\frac{R^2\left(v^2+r^2\right)^2}{X}\left(\frac{1}{\chi^2
\left(\frac{r^2}{v^2}+1\right)}d\phi^2+\frac{\chi^4}{\left(1+
\frac{v^2}{r^2}\right)}d\Omega_3^2\right)\right).
\end{eqnarray}
Note that in these coordinates $v$ and $r$ are not perpendicular.
In the AdS limit $r, v$ become $\rho, u_6$ above though, and hence
the embedding results will be easier to interpret in these
coordinates where we can compare them to the AdS $u_6=m$ result.

From the metric we can calculate the action when we embed with $v$
and $\phi$ as the perpendicular directions. Again we choose
$\phi=const$ and now our variable is $v(r)$.
\begin{eqnarray}
S_{DBI}&=&-T_7\int{d^8\xi}~\frac{e^{4A}R^3\chi^2\sqrt{X}}{v^2+r^2}
\frac{1}{\left(1+\frac{v^2}{r^2}\right)^\frac{3}{2}}\nonumber\\
&&\sqrt{r^2+\frac{R^2 v^2}{\chi^2}+2 r
v(1-\frac{R^2}{\chi^2})v'+(v^2+\frac{R^2 r^2}{\chi^2})v'^2},\nonumber
\label{DBIv}
\end{eqnarray}
where \beq \chi=\chi(\frac{1}{2}\log(v(r)^2+r^2)), \hspace{1cm}
A=A(\frac{1}{2}\log(v(r)^2+r^2)).\eeq

\begin{figure}[!h]
\begin{center}
\includegraphics[height=7cm,clip=true,keepaspectratio=true]{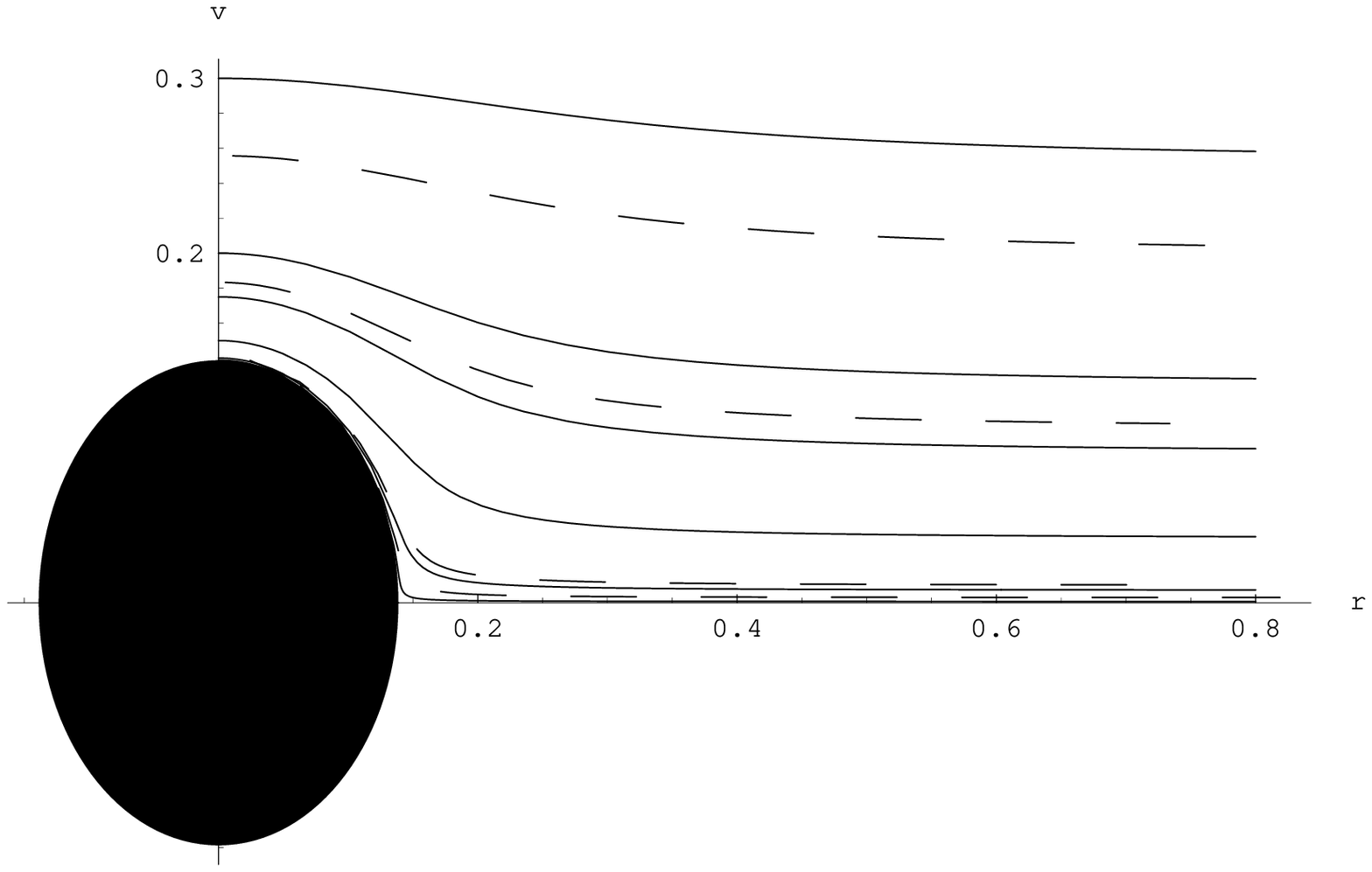}
\caption{D7 brane flow in the FGPW geometry. The solid lines are
the numerical solutions and the dashed lines the coordinate
transform of the full analytic solutions. The singularity of the
geometry is shown as a black circle.}\label{SUSYD7}
\includegraphics[height=7cm,clip=true,keepaspectratio=true]{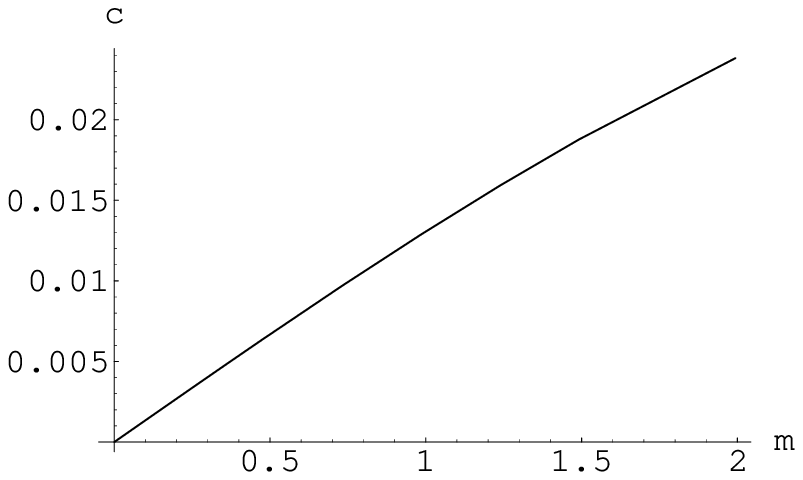}
\caption{Apparent values of the quark mass and condensate
extracted asymptotically from the flows in fig. 1.}\label{wrongmc}
\end{center}
\end{figure}

We can solve the equations of motion resulting from this action
numerically and the solutions are shown in fig.~\ref{SUSYD7}. Note
that the D7 appears to deform around the singularity and there is
a non-zero gradient at large $r$. Naively therefore our embedding
solutions suggest that there is a quark condensate present for
some values of the quark mass - we plot the value extracted
asymptotically also in fig.~\ref{wrongmc}. If there were a quark
condensate for any value of the mass supersymmetry would be broken
which would be a surprise!

In fact, this result is an artifact of being in the wrong
coordinates for the duality to the field theory to be manifest. In
general finding the appropriate coordinates is rather hard but in
this highly supersymmetric theory the correct coordinates have
been identified in \cite{FGPW}\cite{more}. If we change the
coordinates $u, \theta$ for $\hat{u}, \alpha$ so

\beq \hat{u}^2 \cos^2\alpha = R^2 e^{2 A} \rho^2 \cos^2 \theta,
\hspace{1cm} \hat{u}^2 \sin^2 \alpha = R^2 {e^{2A} \over \rho^4}
\sin^2 \theta\label{correctcov},\eeq then the metric takes the
form of a standard multi-centre D3 brane solution:

\beq ds^2 = H^{-1/2} dx_{//}^2 + H^{1/2} \sum_{i=1}^6
d\hat{u}_i^2,\hspace{1cm}  C_{(4)} = {1 \over H g_s} dx^0\wedge dx^1
\wedge dx^2 \wedge dx^3,\eeq with

\beq H = {\chi^2 \over X e^{4A}}. \eeq

The singular region of the space is transformed from a sphere to a
disc by this transformation. The disc lies at the origin in the
$v, r$ space. The singularity is now understood to be caused by
the presence of a disc of D3 branes and the solution is a
perfectly good geometry.

Embedding the D7 brane in the $x_{//}$ and four of the $\hat{w}_i$
directions leaves the standard DBI action (eq.~(\ref{ADSDBI})) for
the D7 in AdS space - the factors of $H$ cancel and play no part.
There are as usual solutions with $\hat{u}_5, \hat{u}_6$ constant
as in AdS. We interpret this as quark fields with a
non-renormalized mass and no quark condensate.

Using the change of coordinates in eq.~(\ref{correctcov}) we can
map these simple $\hat{u}_6$= constant solutions onto the
solutions shown in fig.~\ref{SUSYD7}. These are given by the dashed lines
showing that the numerical solutions to the equations of motion do
match with those acquired from the analytic functions of $\chi$
and A. There are a number of lessons we can learn from this
example. Firstly, it is not straightforward to interpret the D7
embedding solutions in terms of the field theory because of the
potential ambiguity in the identification of the coordinate
system. We can see though there were two signals of the absence of
chiral symmetry breaking in this system even in the wrong
coordinates. Firstly, the value of the parameter $c$ in the
solutions fell to zero as $m$ fell to zero. Secondly, the
embedding solutions filled the full space down to the singularity
- that is there was no clear radial gap between the embedding
solutions and the singularity. The result of this is that, as we
have seen, the singularity can be transformed to a branch cut by a
coordinate transformation and the solution for a massless quark
then lies along the $\hat{u}_6=0$ axis. A true signal of chiral
symmetry breaking, as we'll see shortly, is if there is a radial
gap between the $m=0$ embedding and the singularity - such an
energy scale gap can never be removed by a coordinate
transformation.

\section{Quarks in a Dilaton Flow Geometry}

Chiral symmetry breaking was first observed with a D7 embedding
\cite{Babington:2003vm} in the non-supersymmetric geometry of
Constable and Myers \cite{CM}. This geometry in {\it Einstein
frame} is given by

\beq ds^2 = H^{-1/2} \left( { w^4 + b^4 \over w^4-b^4}
\right)^{\delta/4} dx_{4}^2 + H^{1/2} \left( {w^4 + b^4 \over w^4-
b^4}\right)^{(2-\delta)/4} {w^4 - b^4 \over w^4 } \sum_{i=1}^6
dw_i^2, \eeq where

\beq H =  \left(  { w^4 + b^4 \over w^4 - b^4}\right)^{\delta} - 1,
\eeq and the dilaton and four-form are

\beq e^{2 \Phi} = e^{2 \Phi_0} \left( { w^4 + b^4 \over w^4 - b^4}
\right)^{\Delta}, \hspace{1cm} C_{(4)} = - {1 \over 4} H^{-1} dt
\wedge dx \wedge dy \wedge dz. \eeq There are formally two free
parameters, $R$ and $b$, since \beq \delta = {R^{4} \over 2 b^4},
\hspace{1cm} \Delta^2 = 10 - \delta^2. \eeq

It is convenient to write $w$ and $b$ in units of $R$ which
removes $R$ from the definition of $\delta$ and puts a factor of
$R^2$ into $g_{ww}$. The parameter $b$ has no R-charge and
dimension four so is interpreted as the vev of the operator $\Tr
F^2$. The ${\cal N}=4$ gauge theory is not expected to have a vev
for this operator and hence this geometry probably does not
describe a physical gauge theory vacuum. Nevertheless, it is an
interesting geometry to study chiral symmetry breaking in because
it does describe a non-supersymmetric, strongly coupled gauge
background.

This geometry is particularly simple to study because it has a
flat six plane transverse to the D3s. We parameterize this six
plane as

\beq \sum_{i=1}^6 dw_i^2 = d\rho^2 + \rho^2 d \Omega_3^2 + dw_5^2
+ dw_6^2, \eeq where we will embed the D7 brane on the directions
$x_{//}$, $\rho$ and $\Omega_3$. The U(1) symmetry in the
transverse $w_5-w_6$ plane is the geometric realization of the
$U(1)_A$ symmetry of the quark fields. The DBI action for the D7
is

\beq S=- T_7 \int d^8\xi~
\rho^3\left(\frac{(\rho^2+w_6^2)^4-b^8}{(\rho^2+w_6^2)^4}\right)
\left(\frac{(\rho^2+w_6^2)^2+b^4}{(\rho^2+w_6^2)^2-b^4}\right)^\frac{\Delta}{2}
\sqrt{1+\left(\partial_\rho w_6\right)^2}, \eeq
 and we look for solutions where $w_5=0$ and $w_6$ is
a function of $\rho$. The equation of motion is

\beq \label{eommc}{ d \over d \rho} \left[ {e^{\Phi}  { \cal
G}(\rho,w_6) \over \sqrt{ 1 + \partial_\rho w_6^2}  }
(\partial_\rho w_6)\right] - \sqrt{ 1 + \partial_\rho w_6^2} { d
\over dw_6} \left[ e^{\Phi} { \cal G}(\rho,w_6) \right] =
0, \eeq where

\beq {\cal G}(\rho,w_6) =  \rho^3 {( (\rho^2 + w_6^{2})^2 + b^4) (
(\rho^2 + w_6^{2})^2 - b^4) \over (\rho^2 + w_6^{2})^4}. \eeq The
final terms in the equation of motion is a ``potential" like term
that is evaluated to be

\beq { d \over d w_6} \left[ e^{\Phi} { \cal G}(\rho,w_6) \right]
= { 4 \rho^3 w_6 b^4\over (\rho^2 + w_6^2)^5} \left( {( (\rho^2 +
w_6^2)^2 + b^4) \over ( (\rho^2 + w_6^2)^2 - b^4)}
\right)^{\Delta/2} ( 2b^4  - \Delta (\rho^2 + w_6^2)^2). \eeq
Asymptotically at large radius these equations are just those in
AdS with solution

\beq w_6 = m + { c \over \rho^2} +\dots ,\eeq where $m$ corresponds
to the quark mass and $c$ the condensate.

\begin{figure}[!h]
\begin{center}
\includegraphics[height=5.5cm,clip=true,keepaspectratio=true]{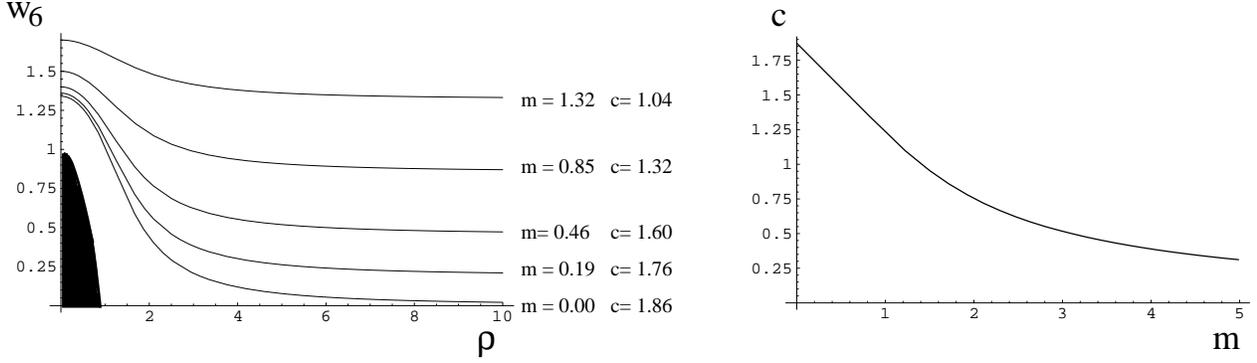}
\caption{Numerical solutions for D7 embeddings in the Constable
Myers geometry with $b=1$. The shaded area corresponds to the
singularity in the metric. Also plotted is the value of the quark
condensate vs the mass extracted asymptotically from those
flows.}\label{CMembed}
\end{center}
\end{figure}

Numerical solutions for the regular embedding solutions are shown
in fig. 3 as well as a plot of the parameter $c$ vs $m$. We take
$b=1$ for these plots. The $m=0$ solution breaks the symmetry in
the $w_5-w_6$ plane which is present asymptotically at large $w$
where the D7 lies at the origin of the space - this is the
geometric breaking of U(1)$_A$.

Let us compare this case to that of the previous section where we
showed that in the ${\cal N}=4$ theory with scalar vev there was
no chiral symmetry breaking. Here there is a non-zero value of $c$
as $m$ goes to zero. There is also a radial gap between the $m=0$
embedding and the singularity. Whatever coordinate transformation
we might make on the geometry this gap will remain and the U(1)
symmetry in the $w_5-w_6$ plane will be broken - this solution
therefore definitely breaks chiral symmetry.

\subsection{Spherical D7 Embedding}\label{spherical}

The above solutions in fig.~\ref{CMembed} are calculated
numerically, but it would be nice to be able to extract the chiral
symmetry breaking behaviour and some dynamical information
analytically. To do this we will look for minimum action solutions
in the form of a circle in the $\rho-w_6$ plane plotted in fig. 3.
The chiral symmetry breaking solutions naively look to match on to
such a solution close to the singularity. Whether such an
embedding falls into the singularity or is stabilized away from it
will test whether the core of the geometry is repulsive to D7s and
hence will be our test for chiral symmetry breaking. The distance
it rests from the singularity will provide an estimate of the
radial gap of the full embedding above. This gap corresponds to
the dynamically generated quark mass at zero energy or the mass
gap of the theory.

Concretely, we write the metric in the six $w$ directions as

\beq \sum_{i=1}^6 dw_i^2 = dr^2 + r^2 (d \alpha^2 + \cos^2 \alpha
d \phi^2 + \sin^2 \alpha d \Omega_3^2), \eeq and embed the D7 brane
on the $\Omega_3$ and in $\alpha$ at constant $r=r_0$. The action
of the D7 brane is

\beq S_{\rm wrapped ~D7} = - T_7\int d^8\xi~ r^4 e^{\Phi(r)}
g_{xx}^2(r)
g_{ww}^2(r)\sqrt{1+\frac{1}{r^2}\left(\frac{dr}{d\alpha}\right)^2}.
\eeq Note that in pure AdS or a multi-centre solution $g_{xx}^2
g_{ww}^2 = 1$, the dilaton is a constant, $d r/d \alpha$ is zero
if $r$ is fixed to some $r_0$ and so the action is simply $r_0^4$.
In the supersymmetric case the circle collapses to the origin
($r_0 \rightarrow 0$) so here the core of the geometry is not
repulsive which we take as evidence of the lack of chiral symmetry
breaking.

\begin{figure}[!h]
\begin{center}
\includegraphics[height=7cm,clip=true,keepaspectratio=true]{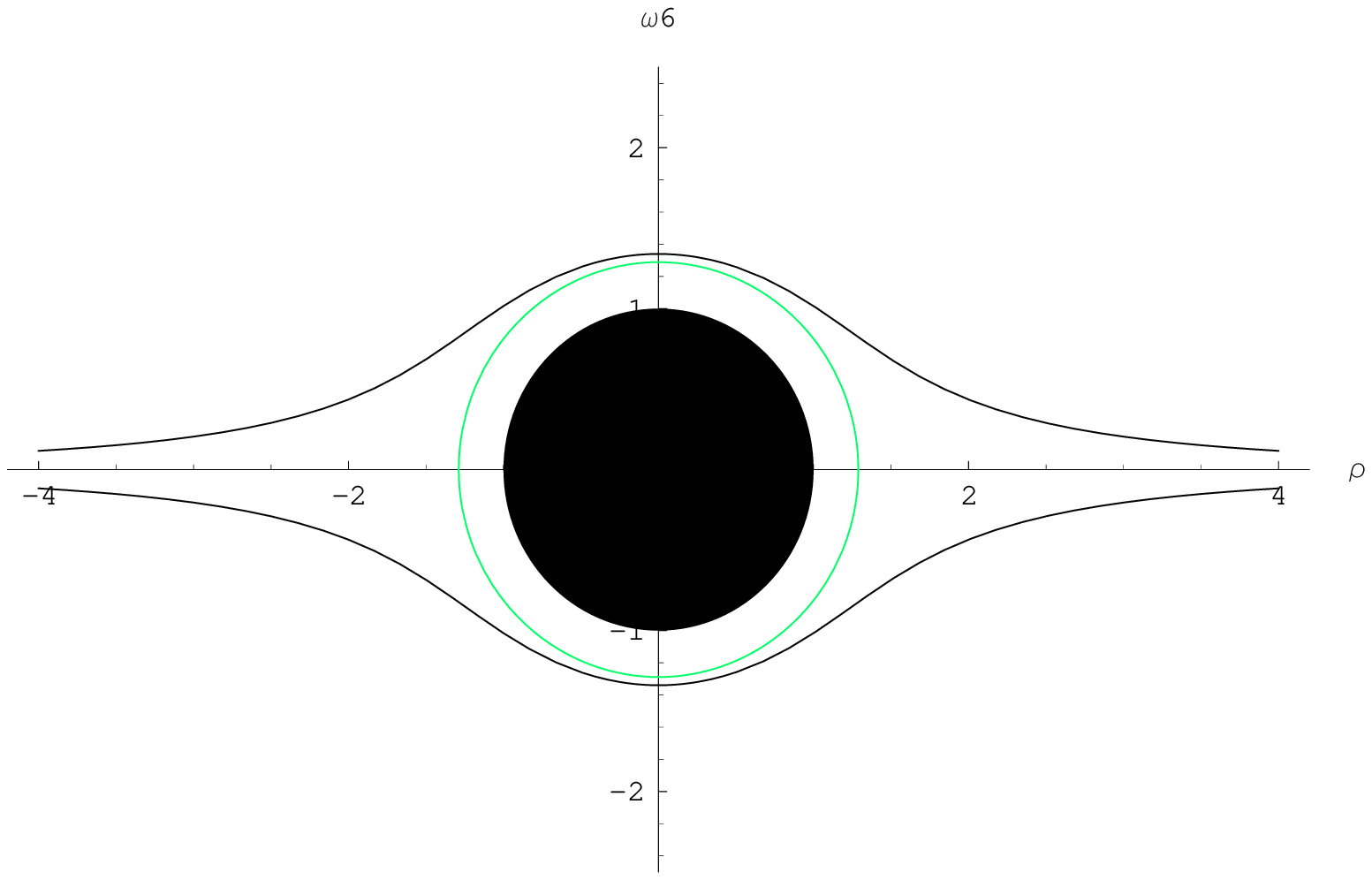}
\caption{Plot of the minimum action spherical D7 embedding and the
massless quark embeddings in the Constable Myers Geometry. The
black circle represents the singularity in the
geometry.}\label{cmmassless}
\includegraphics[height=7cm,clip=true,keepaspectratio=true]{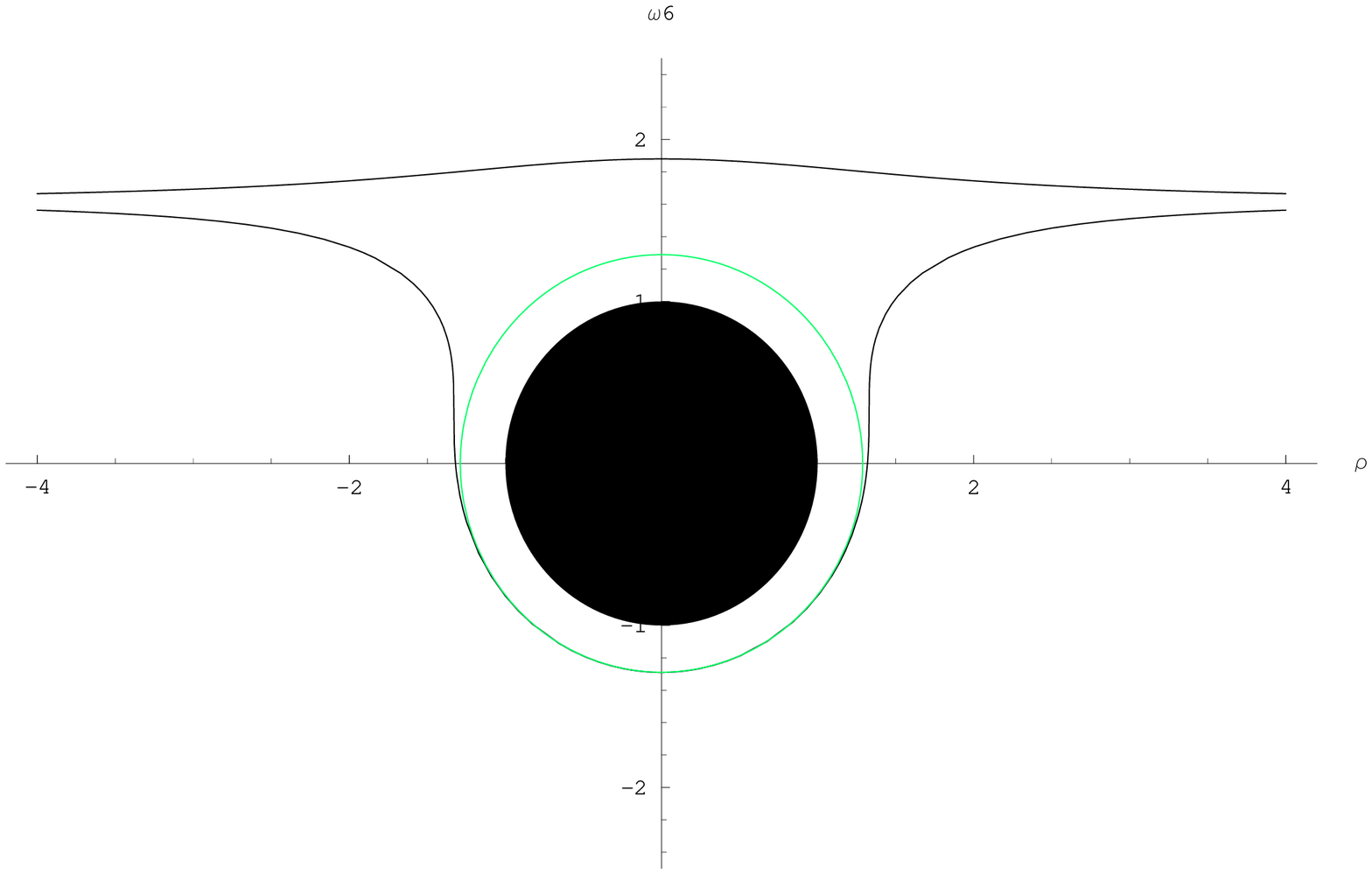}
\caption{Plot of the minimum action spherical D7 embedding and a
local minimum embedding action for a massive quark in the
Constable Myers Geometry. }\label{cmnegcond}
\end{center}
\end{figure}

In the Constable Myers geometry though we have

\beq S_{\rm wrapped ~D7}  =- T_7 \int d^8\xi r^4 \left( {r^4 + b^4
\over r^4 - b^4}\right)^{1 + \Delta/2}  \left( {r^4 - b^4 \over
r^4}
\right)^2\sqrt{1+\frac{1}{r^2}\left(\frac{dr}{d\alpha}\right)^2}.
\eeq Taking the equation of motion for $r(\alpha)$ we see that
there is indeed a solution with $\frac{dr}{d\alpha}=0$. The action
is then minimized by the constant value of $r=r_0$ which is the
root of

\beq \label{quad} r_0^8 - b^4 \Delta r_0^4 + b^8 = 0 \eeq
which is real and greater than b. For $b=1$ there is a minimum of this action away from the
singularity showing that the singularity is repulsive to such a
configuration. For $b=1$  this gives $r_0 = 1.29$. In
fig.~\ref{cmmassless} we plot this solution and the massless quark
solutions above.  Comparison shows that the circular embedding
provides a good approximation to the gap value for the D7 brane
solutions we are interested in above. Generically how good this
match is will depend on the form of the repulsive potential
induced by the geometry. In this case the potential sets in
steeply and the two solutions do match well.

It's clear that the full embedding comes close to matching to the
spherical case in the infra-red. In fact, there is a small gap
between these solutions. In \cite{SE} solutions were identified
that lay closer to the singularity; these correspond to a second,
higher action solution for each asymptotic value of the mass up to
some critical mass. These solutions were identified with the
chiral vacuum with $-\langle \bar{\psi}\psi \rangle$ whose energy
is raised by the quark mass. If the quark mass were too big this
vacuum is not even metastable and there is no solution describing
it. In fig.~\ref{cmnegcond} we plot this critical flow and the
minimum action circular embedding showing that they overlap almost
exactly. Overall though we conclude that the spherical embedding
provides a good measure of the quark mass gap.

There is also some structure as $b$ is changed. Writing $w$ in
powers of $b$ removes $b$ from everywhere in the metric except the
expression for $\Delta$ in the dilaton. For $\Delta$ to be real
requires

\beq b \geq \left( { 1 \over 40}\right)^{1\over 8}R. \eeq

Now the quadratic equation (\ref{quad}) has a solution for the
position of the minimum action, spherically embedded D7 which lies
outside the radius $b$ where the geometry is singular only if

\beq b > \left( { 1 \over 24}\right)^{1\over 8}R. \eeq There is
therefore a small range of $b$ that we can study where the spherical D7 falls into
the singularity. For these values we expect the full D7 embedding
corresponding to the quark fields to fall in too. Numerically this
is what we find. In fact there is a few percent discrepancy
between where a chiral symmetry breaking solution is lost and
where the spherical embedded D7 enters the singularity
corresponding to the slight mismatch seen above for the gap value
in the exact method and the spherical embedding approximation. For
these cases where the singularity is crossed  we can come to no
further understanding without more knowledge of the singularity.

\newpage

\section{A Non-supersymmetric Scalar Deformation}

Let us now turn to our first example of a non-supersymmetric
gravity dual \cite{Babington:2002ci} that has not been previously
studied in this context. The background we will look at is related
to the ${\cal N}=4$ \cite{FGPW} embedding introduced in section 2.
It was generated from a five dimensional supergravity flow which
was lifted to ten dimensions in \cite{Babington:2002ci}. A five
dimensional scalar field, $\lambda$, in the 20 representation of
SO(6) is switched on and has the potential

\beq V=-e^{-\frac{4\lambda(r)}{\sqrt{6}}}-2
e^{\frac{2\lambda(r)}{\sqrt{6}}}. \eeq The scalar field $\lambda$
acts as the source and vev of the field theory operator
$\Tr(\phi_1^2+\phi_2^2+\phi_3^2+\phi_4^2-2\phi_5^2-2\phi_6^2)$.
Switching on a mass term will give rise to unbounded directions in
the scalar potential and so, as with the Constable Myers case,
this is not a realistic dual. Nevertheless it is an interesting
case to see if the breaking of supersymmetry generates chiral
symmetry breaking - we will see that in this case it does not.

\begin{figure}[!h]
\begin{center}
\includegraphics[height=7cm,clip=true,keepaspectratio=true]{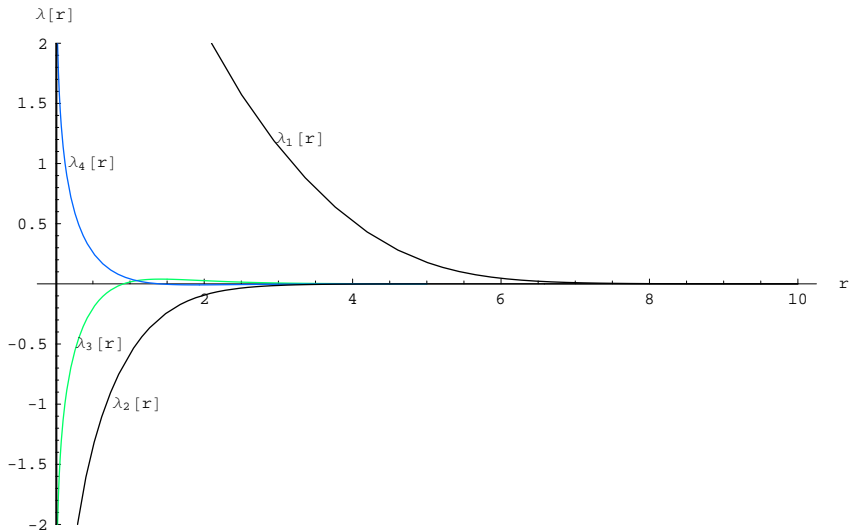}
\caption{4 different flows of the SUGRA field $\lambda(r)$ all of
which become singular at r=0. $\lambda_1$ and $\lambda_2$ are the
turning point flows between the flows of the form $\lambda_3$ and
$\lambda_4$.}\label{lambdas}
\end{center}
\end{figure}

The relevant equations of motion for the scalar field are given by
the usual five dimensional field equations \cite{gppz2}:
\begin{eqnarray}
\lambda''(r)+4\lambda'(r)\sqrt{\frac{1}{6}(\lambda'(r)^2-2 V)}
&=&\frac{\partial V}{\partial\lambda(r)},\nonumber\\
A'(r)&=&\sqrt{\frac{1}{6}(\lambda'(r)^2-2 V)}.
\end{eqnarray}
The large $r$ limit of this field is given by \beq \lambda= {\cal
A} r e^{-2r}+ {\cal B} e^{-2r}\label{lambdaUV}, \eeq where ${\cal
A}$ is interpreted as a source for our operator and ${\cal B}$ is
the vev. The ${\cal N}=4$ deformation described earlier is the
special case where only the vev is non-zero. Now, we are
interested in the case where there is both a mass and condensate
present. We plot numerical solutions of the five dimensional field
equations in fig. \ref{lambdas}. Generically the flows either
diverge in the IR with $\lambda \rightarrow \pm \infty$. There is a boundary
in the ${\cal A}$-${\cal B}$ plane between these two behaviours.
In fig. \ref{lambdas} we show examples of each of these two
behaviours $\lambda_3, \lambda_4$ and the unique flows that lie on
the boundary between these two regimes. $\lambda_1$ describes this
boundary in the positive ${\cal A}$ positive ${\cal B}$ quadrant
whilst $\lambda_2$ describes the boundary in the quadrant where
they are both negative. We will provide analytic forms for the IR
of these flows below.

The ten dimensional lift of the full set of five dimensional
solutions shares the form of the metric with the supersymmetric
geometry so
 \beq
ds^2=\frac{\sqrt{X}}{\chi}e^{2A}dx^2_{//}+\frac{\sqrt{X}}{\chi}
\left(du^2+\frac{R^2}{\chi^2}\left[d\theta^2+
\frac{\sin^2\theta}{X}d\phi^2+\frac{\chi^6
\cos^2\theta}{X}d\Omega_3^2\right]\right), \eeq where the
parameters are defined in exactly the same way as in the
supersymmetric case (see eqn.~(\ref{defineX})) and  \beq
\chi=e^{\frac{\lambda(r)}{\sqrt{6}}}. \eeq Of course the solutions
for $\chi$  and $A$ differ from the supersymmetric case.  The four
form potential of the lift does not match that in the
supersymmetric case but in neither case does it enter the DBI
action of our D7 brane.

We can look for chiral symmetry breaking using both of the
techniques we have seen in previous examples. The first method is
more direct but a little less enlightening;  we probe with a D7
brane by embedding in the $x_{//},r,\Omega_3$ directions (the
angle $\phi$ again provides the U(1)$_A$ symmetry). The D7 action
is given by eq.~(\ref{SUSYDBI}) and we will solve for flows in
from the IR towards the UV with the symmetric boundary condition $v=const$ and calculate numerically the
geometry of the brane. We saw in the supersymmetric example that
we are not in the ``correct'' coordinate system to make the gauge
theory living on the brane manifest and therefore to find the
quark mass and condensate. In that case, we were helped by the
first order equations and could find the correct coordinate
system. In this more general case, where $\chi$ and $A$ are
solutions of second order equations, we have no hint as to how to
find the correct coordinates.

\begin{figure} \begin{center}
\includegraphics[width=0.7\textwidth]{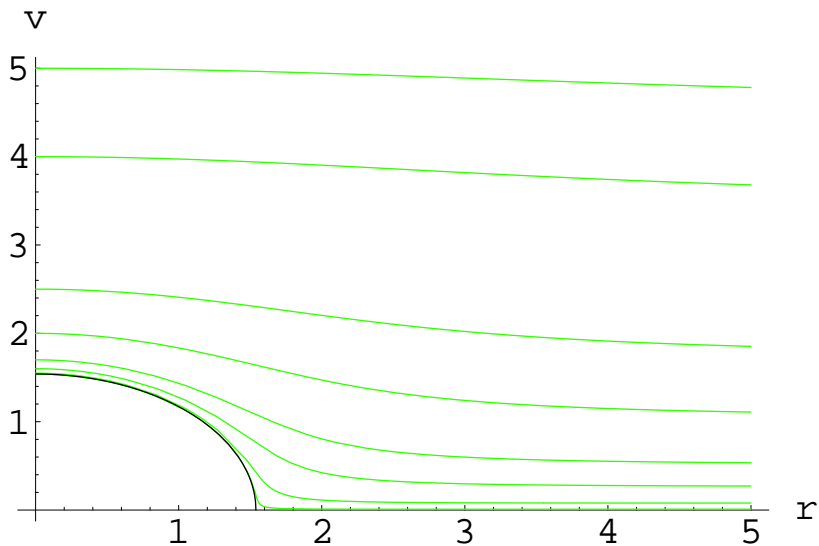}
\caption{Sample solutions for a D7 brane embedding in the
non-supersymmetric scalar deformation geometry showing the absence
of a gap between the solutions and the singularity.} \end{center}
\label{figscaldefcart} \hfill
\end{figure}

We can perform the embedding by calculating $\theta$ as a function
of $r$, or we can change to the cartesian like set  of
coordinates, $(v,r)$,  given by the same change of variables as in
the supersymmetric calculation eq.~(\ref{cov}) and calculate the
flow of $v(r)$. The action is then given in eq.~(\ref{DBIv}). We
solve for the D7 embedding with IR boundary condition $v(r)=$
constant. We plot some representative solutions in
fig.~\ref{figscaldefcart}, which are in a background with positive
${\cal A}$ and ${\cal B}$ in eq.~(\ref{lambdaUV}).

We find generically that for D7 embeddings that correspond to
small quark masses in these geometries the flow wraps onto the
surface of the singularity. As far as we can tell numerically
there is a solution that hugs the $v$ axis down to the singularity
and then follows the singularity to $r=0$. For this flow
asymptotically both the mass and condensate are zero. We conclude
that the geometry does not break chiral symmetries. Note that, as
with the supersymmetric flows, naively a flow that hugs  the
singularity breaks the U(1) symmetry due to $\phi$ translations.
However, it seems likely that there is a coordinate transformation
that presses the singularity to the $v=0$ axis as in the
supersymmetric case. The absence of a gap between the D7 brane
embedding and the singularity, which could not be removed by such
a transformation, implies no clear signal of chiral symmetry
breaking.

The numerical studies above are somewhat messy and it would be
nice to have an analytic understanding of the results. Such
results are provided by our spherical embedding method developed
in section~\ref{spherical}. We study the action of a circular D7
brane wrapped on $x_{//}, \Omega_3, \theta$ at constant $r$ as it
gets nearer to the singular region. We expect the full solutions
above to match onto these solutions in the IR. If the spherical
embedding collapses into the singularity then there will be D7
embeddings that flow in from infinite radius that also touch the
singularity. This can be calculated from the numerical flows of
the scalar field $\lambda$, but it is more satisfactory to be able
to get an analytic form for the potential. It turns out that we
can do this for the IR limit of the scalar flow equations. We find
four different solutions for the scalar fields $\chi$ and $A$.
These numbered solutions to $\chi$ are related to the four types
of behaviour in Fig.~\ref{lambdas} by
$\chi=e^{\frac{\lambda}{\sqrt{6}}}$: \beq
\begin{array}{ccc} \chi_1(r)=\sqrt{\frac{45}{2}}\frac{1}{(r-r_s)},
& \hspace{1cm} &
A_1(r)=4 \log(c(r-r_s)),\\
&&\\
\chi_2(r)=(\frac{2}{3}(r-r_s))^\frac{1}{2}, & \hspace{1cm} &
A_2(r)=\log(c(r-r_s)),\\
&&\\
\chi_3(r)=(b(r-r_s))^\frac{1}{4},& \hspace{1cm} &
A_3(r)=\frac{1}{4}\log(c(r-r_s)),\\
&&\\
\chi_4(r)= (b(r-r_s))^{-\frac{1}{4}}, & \hspace{1cm} &
A_4(r)=\frac{1}{4}\log(c(r-r_s)),\end{array} \label{chis}\eeq
where $r_s$ is the singular radius and $b$ and $c$ are free
parameters \footnote{ We have found this asymptotic behaviour  by
looking for solutions of the form $\lambda(r)=a\log(b(r-r_s))$ as
$r\rightarrow r_s$. In fact, the last two solutions are valid for
any potential V which asymptotically smaller than
$\frac{1}{(r-r_s)^2}$ as $r\rightarrow r_s$. In this case the
supergravity equation reduces to the first order equation: \beq
\lambda''(r)+4\lambda'(r)\sqrt{\frac{1}{6}(\lambda'(r)^2)}\sim
0\nonumber \eeq}. We have checked that these are all of the
possible solutions by numerically solving the flow equations with
these as the IR conditions. We flow out to the UV with these
boundary conditions and calculate the mass and condensate in this
limit. As best we can tell these are indeed all the IR
solutions of the SUGRA field equations.

Having found the analytic IR solutions, we can look at the action
for the circular brane wrapping. We calculate the pullback for a
brane with $r$ and $\phi$ as the perpendicular directions with
$r(\theta)$ and $\phi$ constant which gives: \beq S_{DBI}=-T_7
\int d^8\xi R^4\sqrt{\cos(\theta)^2+\chi(r(\theta))
\sin^2(\theta)}e^{4
A(r(\theta))}\chi(r(\theta))\cos^3(\theta)\sqrt{1+\left(\frac{dr(\theta)}{d\theta}\right)^2
\frac{\chi(r(\theta))^2}{R^2}} \eeq There are solutions where
$r(\theta)=r_0$ and using the above IR solutions we can write down
the action for each of these which gives
\begin{eqnarray}
S_{DBI 1}&\sim& -\int d^8 \xi (r_0-r_s)^{12} |\sin(\theta)||\cos^3(\theta)|,\nonumber\\
S_{DBI 2}&\sim& -\int d^8 \xi (r_0-r_s)^{\frac{9}{2}} \cos^4(\theta),\nonumber\\
S_{DBI 3}&\sim& -\int d^8 \xi (r_0-r_s)^{\frac{5}{4}} \cos^4(\theta),\nonumber\\
S_{DBI 4}&\sim& -\int d^8 \xi  |\sin(\theta)||\cos^3(\theta)|.
\end{eqnarray}
We can see that for the first three there is obviously a
minimizing solution for which $r_0=r_s$ meaning that the circular
brane will fall into the singularity. This strongly suggests from
our earlier studies that there will not be any chiral symmetry
breaking induced by this geometry. Indeed we saw above that
numerical studies of the D7 embedding appropriate for the addition
of quarks also suggest there is not chiral symmetry breaking here.
It is not immediately apparent that the brane will collapse in the
fourth solution, but we note that this solution interpolates from
the UV for which the potential goes like $e^{4r}$. By performing
this calculation numerically we find a monotonically decreasing
potential indicating that the flat behaviour seen in the equation
interpolates smoothly into the exponential behaviour meaning that
the brane collapses with this solution as well.

We conclude that this non-supersymmetric geometry does not induce
chiral symmetry breaking. The unrealistic nature of the geometry,
having as it does an unbounded scalar potential, does not make
this result overly concerning but it is interesting that
supersymmetry breaking and chiral symmetry breaking do not appear
to be necessarily directly linked.

\section{N=2$^*$ Geometry}

We now turn to a more complicated supersymmetric geometry
describing the ${\cal N}=2^*$ gauge theory
\cite{PW}\cite{Brandhuber}. This theory is the ${\cal N}=4$ theory
with mass terms for two adjoint chiral matter fields. It has the
massless fields of ${\cal N}=2$ super Yang Mills and thus a two
dimensional scalar moduli space. The supergravity description has
a field corresponding to the fermionic mass and another which
describes both the scalar masses and vev.  The lift of the
original 5d background to 10d has been made \cite{PW} and the
solution in the Einstein frame is given by \beq
ds^2=\Omega^2\left(e^{2A}dx_{//}^2+dr^2\right)+
\frac{R^2\Omega^2}{\chi^2}\left(\frac{d\theta^2}{c}+\chi^6
\cos^2\theta\left(\frac{\sigma^2_1}{cX_2}+\frac{\sigma_2^2
+\sigma_3^2}{X_1}\right)+\frac{\sin^2\theta}{X_2}d\phi^2\right),
\eeq where
\begin{eqnarray}
\Omega^2&=&\frac{\left(cX_1 X_2\right)^\frac{1}{4}}{\chi},\nonumber\\
c&=&\cosh 2\zeta,\nonumber\\
X_1&=&\cos^2\theta+\chi^6 c \sin^2\theta,\nonumber\\
X_2&=& c \cos^2\theta+\chi^6 \sin^2\theta,\nonumber\\
C_{(4)}&=&\frac{e^{4A}X_1}{4 g_s \chi^2}d x^0\wedge d x^1\wedge d x^2\wedge d x^3,
\end{eqnarray}
and the axion/dilaton is given by: \beq
\lambda=i\left(\frac{1-B}{1+B}\right)=C_{(0)}+i e^{-\Phi},\,~~~
B=e^{2i\phi}\left(\frac{b^\frac{1}{4}-b^{-\frac{1}{4}}}
{b^\frac{1}{4}+b^{-\frac{1}{4}}}\right),\,~~~
b=\frac{X_1}{X_2}\cosh 2\zeta. \eeq In addition, there is the
anti-symmetric two form, whose NSNS and RR parts are given
respectively by
\begin{eqnarray}
B_{(2)} &=& a_3\cos\phi \hspace{2pt}\sigma_1 \wedge \d \phi +
a_1\sin\phi \hspace{2pt}\d \theta \wedge \sigma_1 -
a_2\sin\phi \hspace{2pt}\sigma_2 \wedge \sigma_3,\nonumber \\
C_{(2)} &=& -a_1\cos\phi\hspace{2pt} \d \theta \wedge \sigma_1 +
a_2\cos\phi \hspace{2pt}\sigma_2 \wedge \sigma_3 + a_3\sin\phi
\hspace{2pt}\sigma_1 \wedge \d \phi,
\end{eqnarray}
where
\begin{eqnarray}
a_1 &=& R^2 \tanh 2\zeta \cos \theta,\nonumber \\
a_2 &=& R^2 \frac{\chi^6 \sinh 2\zeta}{X_1}\sin \theta \cos^2 \theta, \nonumber \\
a_3 &=& R^2 \frac{\sinh 2\zeta}{X_2}\sin \theta \cos^2 \theta.
\end{eqnarray}
The SUGRA fields $\zeta$, $A$ and $\chi=e^\alpha$ satisfy the
equations of motion
\begin{eqnarray}
\frac{d\alpha}{d r}&=&\frac{1}{3 R}\left(\frac{1}{\chi^2}-\chi^4 \cosh(2\zeta)\right),\nonumber\\
\frac{d A}{d r}&=&\frac{2}{3 R}\left(\frac{1}{\chi^2}+\frac{1}{2}\chi^4 \cosh(2\zeta)\right),\nonumber\\
\frac{d\zeta}{d r}&=&-\frac{1}{2 R}\chi^4 \sinh(2\zeta).
\end{eqnarray}
These have partial solutions:
\begin{eqnarray}
e^A&=&k\frac{\chi^2}{\sinh(2\zeta)},\nonumber\\
\chi^6&=&\cosh(\zeta)+\sinh^2(2\zeta)\left(\gamma+\log (\tanh\zeta)\right),\label{gammaeq}
\end{eqnarray}

The solutions with large $\gamma$ deform into ${\cal N}=4$
solutions of the form seen in section 2 - here the scalar vev is
so much larger than the supersymmetry breaking mass that the
theory is effectively the ${\cal N}=4$ theory. The smallest
possible vev in the theory corresponds to the background with
$\gamma=0$ and we will concentrate on this case since it is the
vacuum most distinct from the ${\cal N}=4$ theory. Probing the
geometry with a D3 brane \cite{EJP, BPP} shows that the
$\theta=\pi/2$ plane corresponds to the moduli space. Further in
\cite{BPP} a set of coordinates were found on that moduli space
that correspond to the physical coordinates for computing the
$\beta$-function of the field theory. In these coordinates the
singularity is again transformed to a branch cut.

As we have done in previous cases, we can analytically study  the
asymptotic solutions to the flow equations in the IR limit. The
results are\footnote{We find these by looking for real solutions
to the flow equations which behave asymptotically like
$\zeta(r)\sim a \log(b(r-r_s)/R), \chi(r)\sim c ((r-r_s)/R)^d$ as
$r\rightarrow r_s$. It can be shown that these solutions
correspond to $\gamma=0$ as, for $\gamma=0$, we know from
(\ref{gammaeq}) that, for large $\zeta$, $\chi \sim
\left(\frac{4}{3}\right)^\frac{1}{6}e^{-\frac{\zeta}{3}}$, as can
be checked for our solution.}
\begin{eqnarray}
\zeta(r)&=&-\frac{3}{2}\log\left(\left(\frac{2}{3^5}\right)^\frac{1}{3}\frac{(r-r_s)}{R}\right),\nonumber\\
\chi(r)&=&\frac{\sqrt{2}}{3}\sqrt{\frac{(r-r_s)}{R}},\nonumber\\
A(r)&=&4\log\frac{(r-r_s)}{R}+b.
\end{eqnarray}

These are a one parameter family of solutions as expected  due to
the fact that the flow equations in this case are first order. The
parameter $b$ is related to $k$ by $b=\log k
+\log\left(\frac{8}{2187}\right)$. This is not a free parameter,
but is fixed by requiring that this flows to a solution which is
asymptotically $A(r)\rightarrow r$ in the UV.

Now consider including quark fields via a D7 brane probe in this
geometry. The quark superfields have a superpotential coupling to
the ${\cal N}=4$ adjoint scalars of the form $\tilde{Q}A Q$, where
the adjoint field $A$ is represented in the geometry by the two
transverse directions to the D7 brane. Therefore in this geometry
where the ${\cal N}=4$ fields have already been broken into ${\cal
N}=2$ multiplets we must be careful to embed in such a way that we
do not further break supersymmetry. The probe must lie
perpendicular to the $\theta=\pi/2$ plane since that plane
corresponds to the massless scalar fields.  Were the probe at some
angle to that plane the superpotential term would be with fields
in a mix of superfields in terms of the breaking intrinsic in the
geometry. In this configuration though we expect that we should be
able to include arbitrary mass quarks and maintain supersymmetry.

If we were attempting to find the flow of the brane in from the
UV,  we would want to embed it in the $x_{//},r,\alpha,\beta,\psi$
directions and the DBI action of this embedding can be calculated
easily. However, it can be seen that the two perpendicular
directions $\theta$ and $\phi$ will be dependent on both $r$ and
$\psi$, which will make the equations of motion for the brane
computationally very difficult to solve. Instead, we will fall
back on our wrapping technique. We write the DBI action of the
brane in terms of an embedding where the perpendicular directions
$r$ and $\phi$ are functions of $\theta$ and $\psi$ and we take a
constant value of $\phi$. This time by looking at the symmetries
of the metric we know that the minimum action solution will not be
spherical, however we can find out if there is a repulsive
potential on the brane stopping it falling in on the singularity.

To calculate the potential of a D7 brane in this background we
first rewrite the 3 differentials $\sigma_{1,2,3}$ in terms of the
spherical coordinates $\alpha,\beta,\psi$ using
$\sigma_1=\frac{1}{2}(d\alpha + \cos\psi d\beta),
\sigma_2=\frac{1}{2}(\cos\alpha d\psi + \sin\alpha \sin\psi
d\beta), \sigma_3=\frac{1}{2}(-\sin\alpha d\psi + \cos\alpha
\sin\psi d\beta)$. This gives a metric of the form
\begin{eqnarray}
ds^2&=&\Omega^2(e^{2A}dx_{//}^2+dr^2)+\frac{R^2\Omega^2}{\chi^2}
\left(\frac{1}{c}d\theta^2+\frac{\sin^2\theta}{X_2}d\phi^2\right.
\nonumber\\
&&\left.+\frac{\chi^6\cos^2\theta}{4}\left(\frac{d\psi^2}{X_1}
+\frac{d\alpha^2}{cX_2}+d\beta^2\left(\frac{\sin^2\psi}{X_1}
+\frac{\cos^2\psi}{cX_2}\right)+\frac{2d\alpha
d\beta\cos\psi}{cX_2}\right)\right).
\end{eqnarray}

In fact even this configuration is hard to study because of the
forms present in the geometry. For simplicity we will place the D7
probe at $\phi=n\pi$ where the axion vanishes, and the dilation is
given by \beq e^{\Phi}=\sqrt{\frac{cX_1}{X_2}}.\eeq In addition,
the NSNS two form is zero.

The DBI action for this configuration is given by
\begin{eqnarray}
S_{DBI}&=& -T_7\int d^8\xi \frac{R^4}{4} e^{4A} c X_1\cos^2\theta
|\sin\psi|\nonumber\\
&&\sqrt{\left(\frac{\chi^2\cos^2\theta}{4 cX_1}
+\frac{\chi^4\cos^2\theta}{4 R^2 X_1}\left(\frac{\partial r}{\partial\theta}\right)^2
+\frac{1}{c R^2\chi^2}\left(\frac{\partial r}{\partial\psi}\right)^2\right)}.
\end{eqnarray}

We also have to consider the Wess-Zumino part of the action for
the D7 brane.  As there are no gauge fields living on the brane,
this is given by \beq
S_{WZ}=\mu_7\int_{\mathcal{M}_8}(C_{(8)}+C_{(6)}\wedge
B_{(2)}).\label{WZaction} \eeq For $\phi=n\pi$ this will be zero as
the dual of $C_{(8)}$ (the axion) is zero and $B_{(2)}$ is also zero.

As before, we are only interested in the IR behaviour of the
potential  felt by the brane so we can use our analytic solution
in the above equation. Ignoring proportionality factors, we find
that for a constant $r_0$ solution \beq V_{IR}\sim
(r_0-r_s)^{15}|\cos^3\theta||\sin\psi|\sqrt{5+\cos 2\theta}. \eeq
We can see clearly that, as expected, this supersymmetric
background, when probed in this particular way, does not have the
signature of chiral symmetry breaking. There is no potential
stopping the spherical D7 brane falling onto the singularity and
the situation is analagous to that we found in the ${\cal N}=4$
geometries.\newpage

\section{The Yang Mills$^*$ Geometry}

Finally we will turn our spherical D7 embedding technique on a
geometry where physically one might expect chiral symmetry
breaking. The \yms geometry \cite{Babington:2002qt} was developed
as a model of non-supersymmetric Yang Mills theory. The UV of the
theory is the ${\cal N}=4$ gauge theory but then at a scale ${\cal
M}$ a mass term is introduced for the four adjoint fermions,
$\lambda_i$. It is also possible to include in the solutions a vev
for the operator $\sum_i \lambda_i \lambda_i$. One would expect
some dynamical determination of this condensate in terms of ${\cal
M}$ but the supergravity solution does not clearly provide this
link - we will investigate this whole space of geometries
therefore.

The \yms geometry was originally constructed as a 5d supergravity
solution but then lifted to a full 10d solution. In 10d D3 brane
probe analysis indicates that the six adjoint scalars of the
${\cal N}=4$ theory acquire masses radiatively as one would expect
since supersymmetry is broken. In \cite{Apreda:2003} the glueball
spectrum and string tension properties were analyzed. For the
geometries with a fermion mass and small or vanishing condensate,
a discrete glueball spectrum was found though probe strings fell
onto the singularity. The interior of the geometry is therefore
still ill understood, and possibly is also incomplete since the
restricted 5d solution on which the geometry is built may not have
sufficient freedom to describe the full non-supersymmetric theory.
Nevertheless, this is a prime candidate geometry to examine for
chiral symmetry breaking since it describes a model that shows
some very QCD like qualities.

In the SUGRA theory the fermion mass terms correspond to turning
on a scalar in the 10 of SO(6). As in the other examples, we can
solve the 5d SUGRA equations numerically using the relevant field
equations.
\begin{eqnarray}
\lambda''(r)+4A'(r)\lambda'(r)=\frac{\partial V}{\partial\lambda},\nonumber\\
A'(r)=\sqrt{\frac{1}{6}(\lambda'(r)^2-2V)},
\end{eqnarray}
with \beq V=-\frac{3}{2}(1+\cosh(\lambda(r))^2). \eeq The AdS
limit of this field has solutions \beq \lambda(r)={\cal M} e^{-r}+
{\cal C}e^{-3r}, \eeq with ${\cal M}$ and ${\cal C}$, the mass and
condensate of the operator $\sum_i \lambda_i \lambda_i$
respectively. We will be particularly interested in the large
negative $r$ limit of the space corresponding to the IR of the
gauge theory. We can try to find analytic solutions in this limit
as we did in the previous section. The solutions we have found are

\beq \begin{array}{ll}
\lambda_{IR,1}(r)=-\log\left(\sqrt\frac{9}{20}(r-r_s)\right),&
\hspace{1cm} A(r)=\frac{2}{3}\log (r-r_s),  \\ & \\
\lambda_{IR,2}(r)=\log\left(\sqrt\frac{9}{20}(r-r_s)\right),&
\hspace{1cm} A(r)=\frac{2}{3}\log (r-r_s),\\ & \\
\lambda_{IR,3}(r)=-\frac{\sqrt{6}}{4}\log(a(r-r_s)),& \hspace{1cm}
A(r)=\frac{1}{4}\log (r-r_s),\\& \\
\lambda_{IR,4}(r)=+\frac{\sqrt{6}}{4}\log(a(r-r_s)), &
\hspace{1cm} A(r)=\frac{1}{4}\log (r-r_s),\end{array} \eeq
\bigskip

\noindent where $r_s$ is the radius where the flow becomes
singular in the IR. In the ${\cal C}$ vs ${\cal M}$ plane there
are regions where the $\lambda$ flows diverge positively and
negatively described by the second two solutions. There is then a
unique flow on the boundary which in the positive quadrant is
described by $\lambda_1$ and in the negative quadrant by
$\lambda_2$. Numerically the unique flow lies at least close to
the boundary condition ${\cal C}=0$.

The 10d lift of the \yms background is given by
\begin{eqnarray}
ds^2_{10}&=&\xi^\frac{1}{2}(e^{2A}dx^2+dr^2)
+\xi^{-\frac{3}{2}}(\xi^2 d\alpha^2+\sin^2\alpha F_+d
\Omega_+^2+\cos^2\alpha F_-d\Omega_-^2),\nonumber\\
d\Omega^2_\pm&=&d\theta^2_\pm+\sin^2\theta_\pm d\phi^2_\pm,\nonumber\\
\xi^2&=&\cosh^4\lambda(r)+\sinh^4\lambda(r)\cos^2 2\alpha,\nonumber\\
F_\pm&=&\cosh^2\lambda(r)\pm \sinh^2\lambda(r)\cos 2\alpha,\nonumber\\
A\pm&=&\frac{\sinh 2\lambda(r)}{\cosh^2\lambda(r)\pm\cos 2\alpha \sinh^2\lambda(r)},\nonumber\\
B&=&\frac{\sinh^2\lambda(r) \cos2\alpha}{\cosh^2\lambda(r)+\xi},\nonumber\\
e^{-\Phi}&=&\frac{1-B}{1+B}.\label{dil}
\end{eqnarray}

Now consider including quark fields via a D7 brane probe in this
geometry. The probe would naturally lie in the $x_{//}$ and $r$
directions and then wrap a three sphere in the deformed five
sphere. For example we could wrap one of the two spheres (e.g.
$\Omega_-$) and fill the angle $\alpha$ leaving us to find the
embedding $\theta_+$. The angle $\phi_+$ provides the U(1)$_A$
symmetry of the quarks. Clearly $\theta_+$ will be a function of
both $r$ and $\alpha$ on the D7 world volume. In this complicated
geometry it is too difficult to find the full embedding \footnote{
Note that the usual AdS geometry can be written in the same
coordinate system as \yms so the metric is \beq ds^2 = e^{2r}
dx_{//}^2 + dr^2 + d\alpha^2 + \cos \alpha d\Omega_+^2 + \sin
\alpha d \Omega_-^2 \eeq The flat D7 embedding of section 2 in
these coordinates is then given by \beq \theta_- = \arcsin { m
e^{-r} \over \sin \alpha}\eeq which is itself a complicated
function. Even in this case sophisticated numerical techniques
would be needed to find this solution.}. It is more
straightforward though to embed a D7 spherically on $x_{//},
\Omega_-, \alpha$ and $\theta_+$ at fixed $\phi_+$. We treat the
radius of this sphere $r$ as our embedding coordinate. The DBI
action for this configuration is
 \beq
S_{DBI}=-T_{D7}\int d^8\xi~
e^\Phi\sqrt{-M}\left((\partial_{\theta_{+}}r)^2+\left(1+(\partial_\alpha
r)^2\right)\frac{\sin^2\alpha}{F_-}\right)^\frac{1}{2}
,\eeq

where
\beq
\sqrt{-M}=e^{4A}\sqrt{\cos^4\alpha F^2_- \sin^2{\theta_-}+\frac{A^2_-}{4}\sin^6\alpha \cos^2\theta_-\xi^3}.
\eeq

The Wess-Zumino contribution eq.(\ref{WZaction}) of the action
must also be studied. However, the integral over $C_{(8)}$ is zero
as it's dual, the axion, is zero and the integral over $C_{(6)}
\wedge B_{(2)}$ will also be zero since $C_{(6)}$ is dual to
$C_{(2)}$ ($\d C_{(6)}= \hspace{1pt}*{\d C_{(2)}}$) and $C_{(2)}$
has a basis $\d\theta_+ \wedge \d\phi_+$, whereas $B_{(2)}$ has a
basis $\d\theta_- \wedge \d\phi_-$. This means that we will end up
with a wedge product of identical basis one forms which will be
zero. For this embedding we can therefore drop these terms.

The symmetries of our action imply that there are solutions where
$r$ is just a function of $\alpha$ which is to be compared with
the more complicated full embedding where $\theta_+(r, \alpha)$.
Numerical analysis will therefore be simpler in this case.

As our variable is not dependent on $\theta_-$, we must integrate
over this quantity in the action before we can try to find a
solution. This integral gives \beq S_{DBI}=-T_7\int d^7\xi~
e^\Phi e^{4A}2F_{-}\cos^2\alpha ~
EllipticE\left(1-\frac{A_{-}^2\sin^6\alpha\zeta^3}{4
F_{-}^{2}\cos^4\alpha}\right)\sqrt{(\partial_{\theta_{+}}r)^2+(1+(\partial_\alpha
r)^2)\frac{\sin^2\alpha}{F_{-}}}. \eeq

The simplest analysis we can perform is just to look at the
potential felt by a brane of fixed radius $r_0$. This will not be
a solution of the equations of motion but will show us whether
there is a repulsion in the geometry to such a configuration
collapsing.  For our four analytic solutions, we find two sorts of
behaviour in the IR. These are given by the potentials
 \beq V_{circle}=T_7\int d^7\xi~
(r_0-r_s)^{\frac{2}{3}}~EllipticE\left(1-\frac{18}{5}(r_0-r_s)
^2|\cot\alpha|\right)\sqrt{|\cos^5\alpha\sin^3\alpha|} ,\eeq
corresponding to $\lambda_{IR,1}$ and $\lambda_{IR,2}$ and \beq
V_{circle}=T_7\int d^7\xi~(b
(r_0-r_s))^{1-\sqrt{\frac{3}{2}}}~EllipticE\left(1-8(b
(r_0-r_s))^{\sqrt{\frac{3}{2}}}|\cot\alpha|\right)\sqrt{|\cos^5\alpha\sin^3\alpha|}
,\eeq corresponding to $\lambda_{IR,3}$ and $\lambda_{IR,4}$.  For
both of these solutions the elliptic integral tends to a constant
in the IR so we can see explicitly the behaviour of the potential
as a function of $r_0-r_s$. In the first case, corresponding to a
line of solutions in the ${\cal M}$ vs ${\cal C}$ plane the
embedding collapses onto the singularity. For the remainder of the
${\cal M}$ vs ${\cal C}$ space the probe is repelled from the
singularity in the IR. In the UV where the geometry returns to AdS
the potential always pushes the field into the IR so  there must
be a stable configuration away from the singularity. This suggests
that the majority of parameter space in the model will give rise
to chiral symmetry breaking.

\begin{figure}[!h]
\begin{center}
\includegraphics[height=7cm,clip=true,keepaspectratio=true]{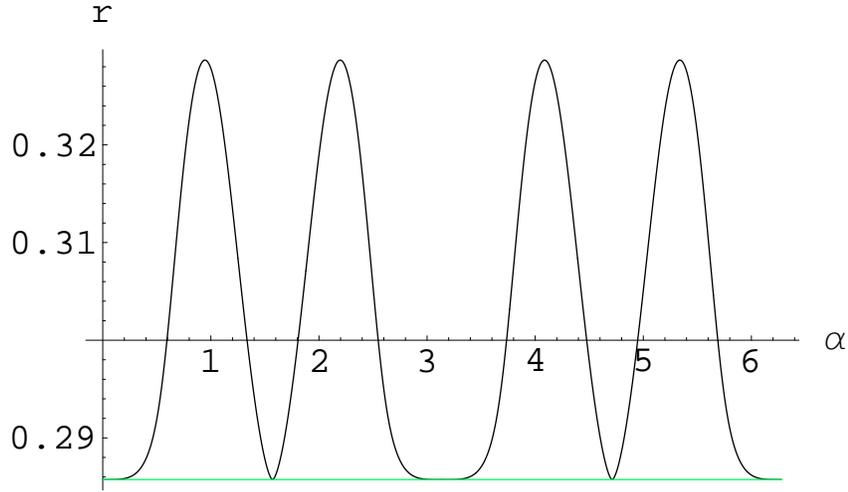}
\caption{Sample result of a numerical solution for the
determination of a spherical D7 brane embedding $r$ vs $\alpha$
that wraps the \yms singularity shown by the lower straight
line.}\label{YMcircle}
\end{center}
\end{figure}

\begin{figure}[!h]
\begin{center}
\includegraphics[height=7cm,clip=true,keepaspectratio=true]{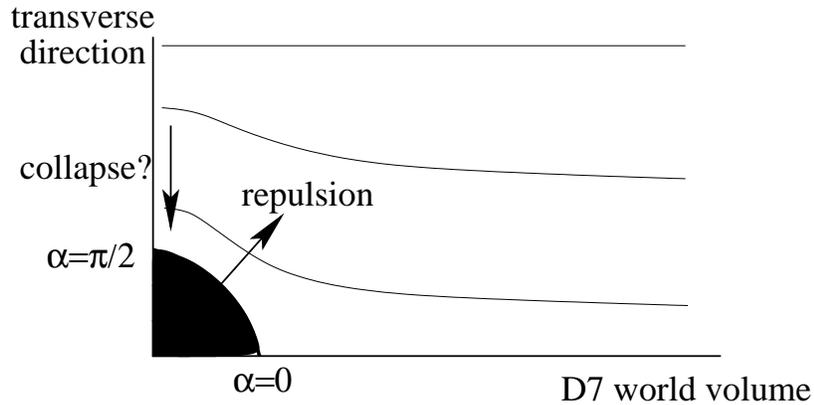}
\caption{A sketch of how a flat D7 brane embedding is expected to
behave in \yms as the brane is brought close to the singularity.
The repulsive potential away from $\alpha=n \pi/2$ will induce a
chiral symmetry breaking like configuration although the D7 may
collapse into the singularity at $\alpha=\pi/2$. }\label{sketch}
\end{center}
\end{figure}

We know however from the $\alpha$ dependence of the action that
the solution of the wrapped brane will not have a circular
symmetry. Therefore, although we know that there is a repulsive
potential at least somewhere around the singularity, we don't yet
know what form the brane embedding will take. To calculate this,
we have used a numerical relaxation method. We discretise the
points on the brane parameterised by the angle $\alpha$ and write
the action for $r(\alpha)$ in the discretised form with a starting
guess for the solution. We then minimise the action with respect
to all points on the lattice, giving us N coupled equations for N
lattice points. Because we are looking at a wrapping solution, we
also have a boundary condition that the first and last points in
$\alpha$ are at the same value of  $r$. We performed this
calculation using Mathematica. The resulting configuration of the
brane is shown in fig. \ref{YMcircle} for a generic flow of the
type we have seen with a repulsive potential. We see that the
repulsive potential is present away from $\alpha=n \pi/2$ but
vanishes at precisely $n \pi/2$ at which points the brane can fall
in on the singularity. This is obviously an added complication
which muddies the result. Consider beginning with a flat D7 brane
far from the singularity which would include a heavy quark field
in the theory. As the brane is brought in towards the singularity
the repulsion centred at $\alpha=\pi/4$ will stop the brane moving
to the origin of the space as sketched in fig \ref{sketch}
implying a chiral symmetry breaking set of UV boundary conditions.
The outstanding question is whether the probe will touch the
singularity at $\alpha=\pi/2$. We can not address this completely
within our analysis. Of course touching the singularity may not be
a disaster if it is simply indicating the presence of a fuzzy
configuration of D3s - some brane bound state may form there. We
tentatively conclude that \yms generically does break chiral
symmetries although there are outstanding issues to be understood.

\section{Summary}

Quark fields can be introduced into the AdS/CFT Correspondence and
its deformations via D7 brane probes. The mass and any chiral
condensate induced can be read off from the asymptotic behaviour
of the scalar in the D7 DBI action describing its embedding. We
have shown that in fact this prescription must be tempered by the
possibility of making coordinate transformations that alter the
asymptotic behaviour of the field. In the FGPW geometry
\cite{FGPW} that describes the ${\cal N}=4$ gauge theory on its
moduli space we saw that the D7 embedding flows fill the whole
space away from the central singularity of the geometry. A
transformation that takes the singularity to a branch cut would
then remove the signal for chiral symmetry breaking. In this case
that is precisely the coordinate transformation that has been
previously identified as necessary in the literature. In contrast,
in the non-supersymmetric dilaton flow \cite{CM} in which chiral
symmetry breaking has previously been studied
\cite{Babington:2003vm}, there are D7 embeddings for all possible
quark masses that lie separated from the core singularity. Here no
coordinate transformation can remove the symmetry breaking
embedding.

To look at more complicated metrics where the full D7 embedding is
numerically too involved to find, we developed a simple spherical
D7 embedding in the dilaton driven flow. The full embedding
appropriate for quark fields matches on to this spherical
embedding in the IR. In the dilaton flow case, we could
analytically find the potential for the spherical embedding and
use it to show that the core of the geometry is repulsive. This
repulsion drives the D7 brane into a chiral symmetry breaking
configuration. We could also compute an analytic estimate of the
quark mass gap.

We then used this spherical embedding technique to test three new
geometries for chiral symmetry breaking behaviour. The first is a
non-supersymmetric version of the FGPW background
\cite{Babington:2002ci} in which an unbounded scalar mass is
included in the gauge theory. D7 branes in this geometry behaved
precisely as those in the supersymmetric version of the  FGPW
geometry leaving no gap between the flows and the singularity. The
spherical embedding technique showed there was no potential
stopping the spherical D7 falling onto the singularity. We
conclude that this geometry does not break chiral symmetries.

We next looked at a spherical D7 embedding in the ${\cal N}=2^*$
geometry \cite{PW} and found again that the D7 falls in on the
singularity. This implies that for a supersymmetric embedding of a
D7 for the inclusion of quarks in the geometry there would be no
chiral symmetry breaking as we would expect.

Finally we studied a spherical D7 embedding in the
non-supersymmetric Yang Mills$^*$ geometry
\cite{Babington:2002qt}. For most of the parameter space we found
the central singularity of the geometry displayed a repulsive
potential. This would imply chiral symmetry breaking embeddings
for quark fields. There are special angles at which the repulsion
vanishes but most probably this will not effect the conclusion.

Interestingly we have found chiral symmetry breaking in only those
theories that are both non-supersymmetric and have a running
dilaton. The former matches our expectations but the generality of
the latter point is unclear. We believe that our spherical
embedding technique will provide a simple test for chiral symmetry
breaking in the more complicated geometries that will be needed to
describe QCD realistically.

\vspace{2.0cm}

\noindent {\bf Acknowledgements}

JS and TW are grateful to PPARC for the support of their
studentships.  

\end{document}